\documentclass[preprint,12pt,1p]{elsarticle}

\usepackage{epsfig}
\usepackage{amssymb}

\journal{Nuclear Physics A}

\begin{document}
\begin{frontmatter}

\title{Jet Quenching with Parton Evolution}

\author[rvt,focal,els]{Luan Cheng\corref{cor1}}
\author[focal,els]{Enke Wang}

\cortext[cor1]{chengluan@iopp.ccnu.edu.cn}

\address[rvt]{College of Mathmatics and Statistics, Huazhong Normal University,\\ Wuhan 430079, China}
\address[focal]{Institute of Particle Physics, Huazhong Normal University, Wuhan 430079, China}
\address[els]{Key Laboratory of Quark $\&$ Lepton Physics, Ministy of Education, China}

\begin{abstract}
We report the evolution effects on jet energy loss with detailed
balance. The initial conditions and parton evolution based on
perturbative QCD in the chemical non-equilibrated medium and Bjorken
expanding medium at RHIC are determined. The parton evolution affect
the jet energy loss evidently. This will increase the energy and
propagating distance dependence of the parton energy loss and will
affect the shape of suppression of moderately high $P_{T}$ hadron
spectra.

\end{abstract}

\end{frontmatter}

One of the challenging goals of heavy-ion physics is to detect
quark-gluon plasma (QGP). Jet quenching \cite{WG1992} or suppression
of large $p_T$ hadrons, caused by the energy loss of a propagating
parton in a dense medium, has become a powerful tool for the study
of properties of QGP. In the heavy ion collisions, the two nuclei
pass through each other, interact, and then produce a dense plasma
of quarks and gluons. As the initial parton density is large and the
partons suffer many collisions in a very short time, the initial
partonic system may attain kinetic equilibrium. But does it attain
chemical equilibrium? From the numerical studies of parton cascade
model, which is based on the concept of inside-outside cascade
\cite{Anishetty:1980sf,Hwa:1986sf,Balizt:1987} and evolve parton
distributions by Monte-Carlo simulation of a relativistic transport
equation involving lowest order perturbative QCD scattering and
parton fragmentations, it is believed that QGP likely to be formed
in such collisions are far from chemical equilibrium. So that the
effect of parton chemical equilibration on jet energy loss need to
be studied.

We consider here a thermal equilibrated, but chemical
non-equilibrated system, and write the parton distribution as an
approximation in the factorized Bose or Fermi-Dirac form with
non-equilibrium fugacities $\lambda_i$ which gives the measure of
derivation from chemical equilibrium,
\begin{equation}
f(k;T,\lambda_i) = \lambda_i(e^{\beta u\cdot k}\pm
1)^{-1}\label{two}\, .
\end{equation}

In general, chemical reactions among partons can be quite
complicated because of the possibility of initial and final-state
gluon radiations. Here we are interested in understanding the basic
mechanisms, so we restrict our consideration to the dominant
reaction mechanisms $gg\leftrightarrow ggg$, $gg\leftrightarrow
q\bar{q}$ for the equilibration of each parton
flavor\cite{Biro:1994sf}. Restricting to reactions, the evolution of
the parton densities is governed by the master equations,
\begin{eqnarray}
\partial_\mu(\rho_{g}u^{\mu})&=&\rho_{g}R_3(1-\lambda_{g})
-2\rho_{g}R_2(1-\frac{\lambda_{q}\lambda_{\bar{q}}}
{\lambda_{g}^2})\, ,
\label{master1}\\
\partial_\mu(\rho_{q}u^{\mu})&=&
\rho_{q}R_2(1-\frac{\lambda_{q} \lambda_{\bar{q}}}
{\lambda_{g}^2})\, ,
\label{master2}
\end{eqnarray}
where $R_2=\frac{1}{2}\sigma_{2}n_g$,
$R_3=\frac{1}{2}\sigma_{3}n_g$, $\sigma_2$ and $\sigma_3$ are
thermally averaged velocity weighted cross sections,
$\sigma_2=\langle\sigma(gg\rightarrow q\bar{q})v\rangle$,
 $\sigma_3=\langle\sigma(gg\rightarrow ggg)v\rangle$
\cite{Biro:1994sf,Wang:1997sf}.

If we assume that parton scatterings are sufficiently rapid to
maintain local thermal equilibrium, and therefore we can neglect
effects of viscosity due to elastic and inelastic scatterings, we
can have the hydrodynamic equation, $
\partial_{\mu}(\varepsilon u^{\mu})+P\partial_{\mu}u^{\mu}=0 \label{hydro1}\, ,
$ and the baryon number conservation $\partial_{\mu}(\rho_{_B}
u^{\mu})=0$. Then with the master equations above, if the four
equations can be solved, we can determine the evolution of
$T(\tau)$, $\lambda_g(\tau)$ and $\lambda_q(\tau)$ towards chemical
equilibrium, once initial conditions are known. So the input of the
initial condition play an important role to investigate the effects
of the evolution for the parton system.

\begin{figure}
\vspace{-20pt}
\includegraphics[width=4.4cm]{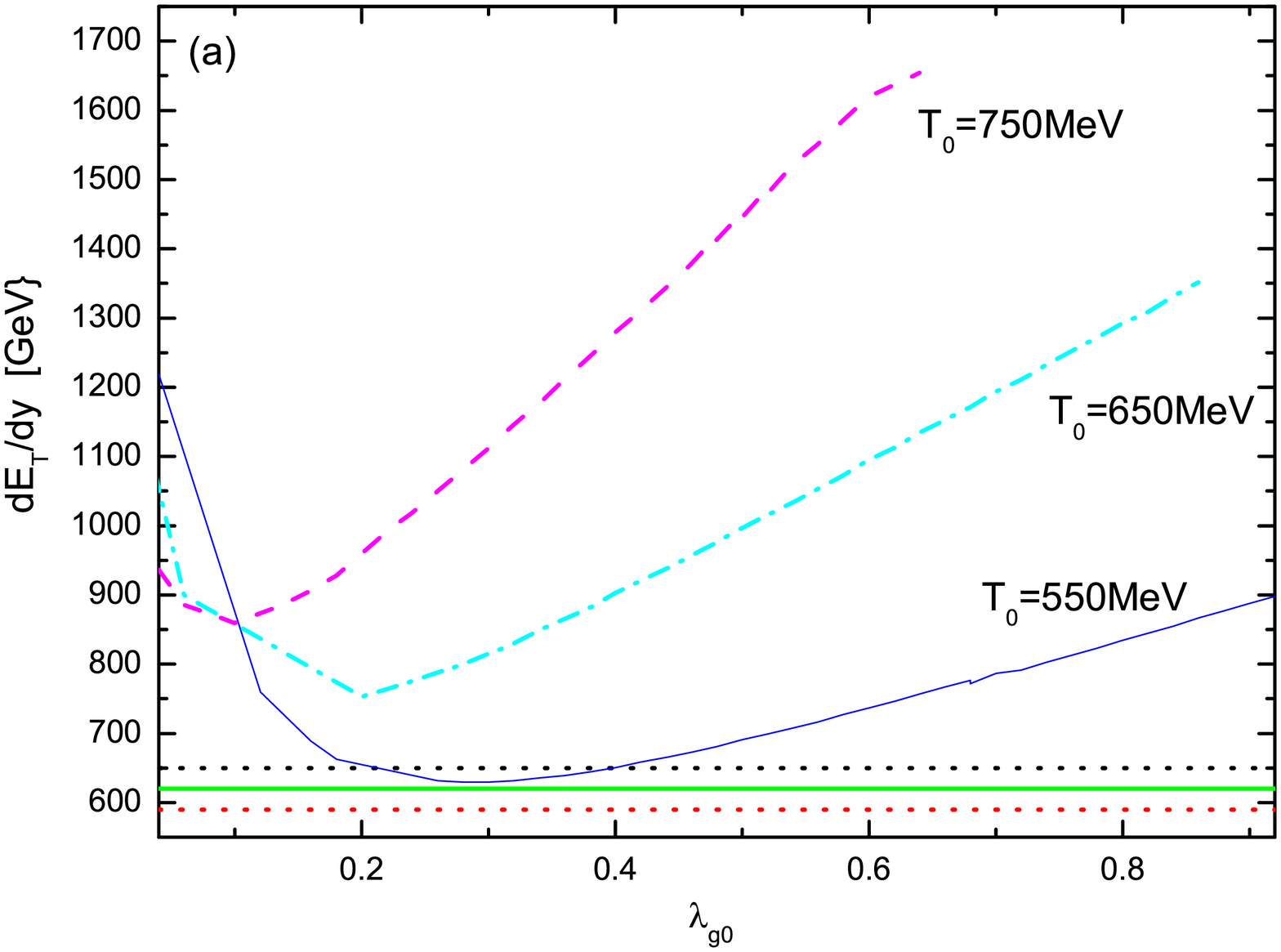}
\includegraphics[width=4.4cm]{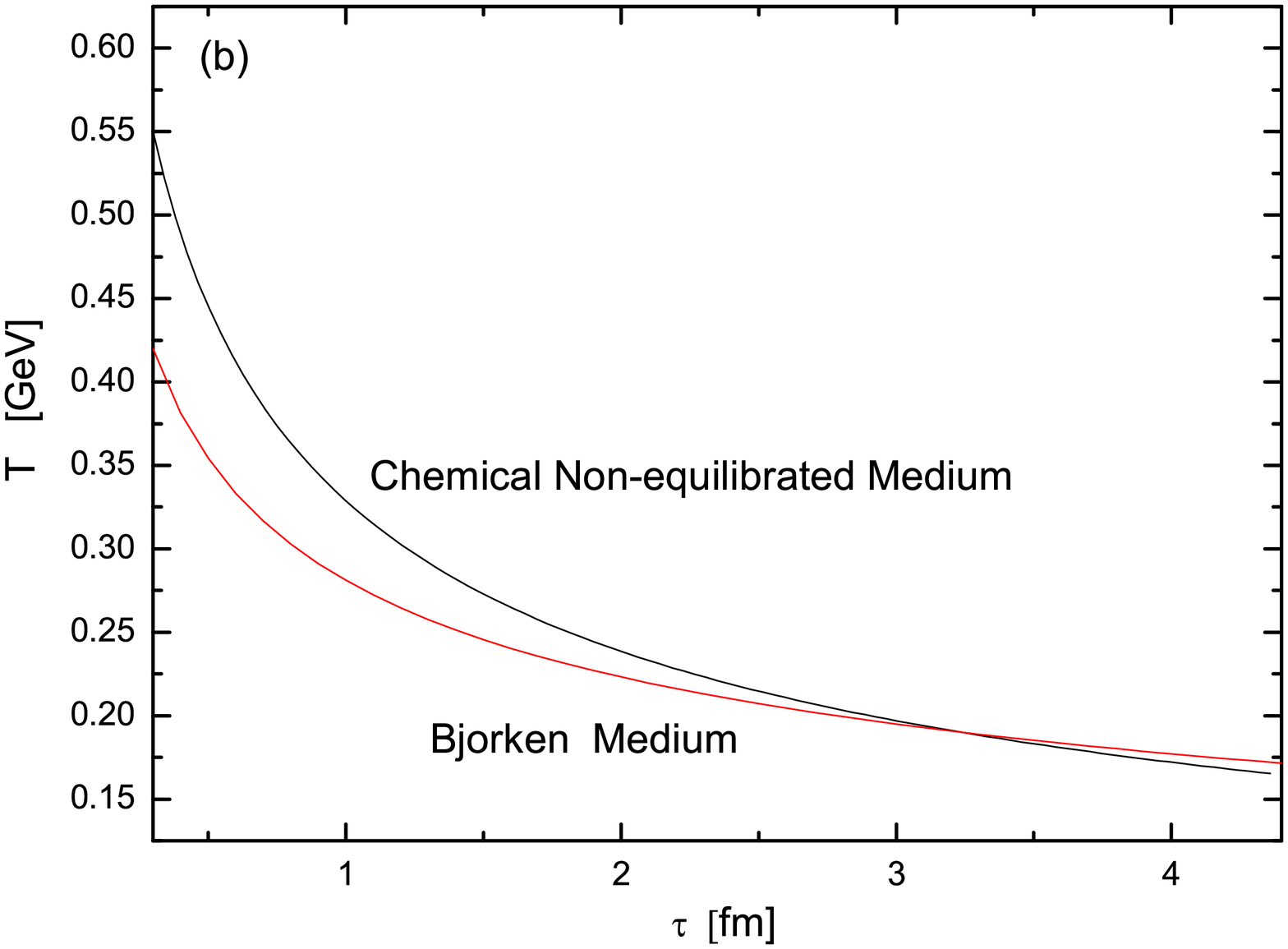}
\includegraphics[width=4.4cm]{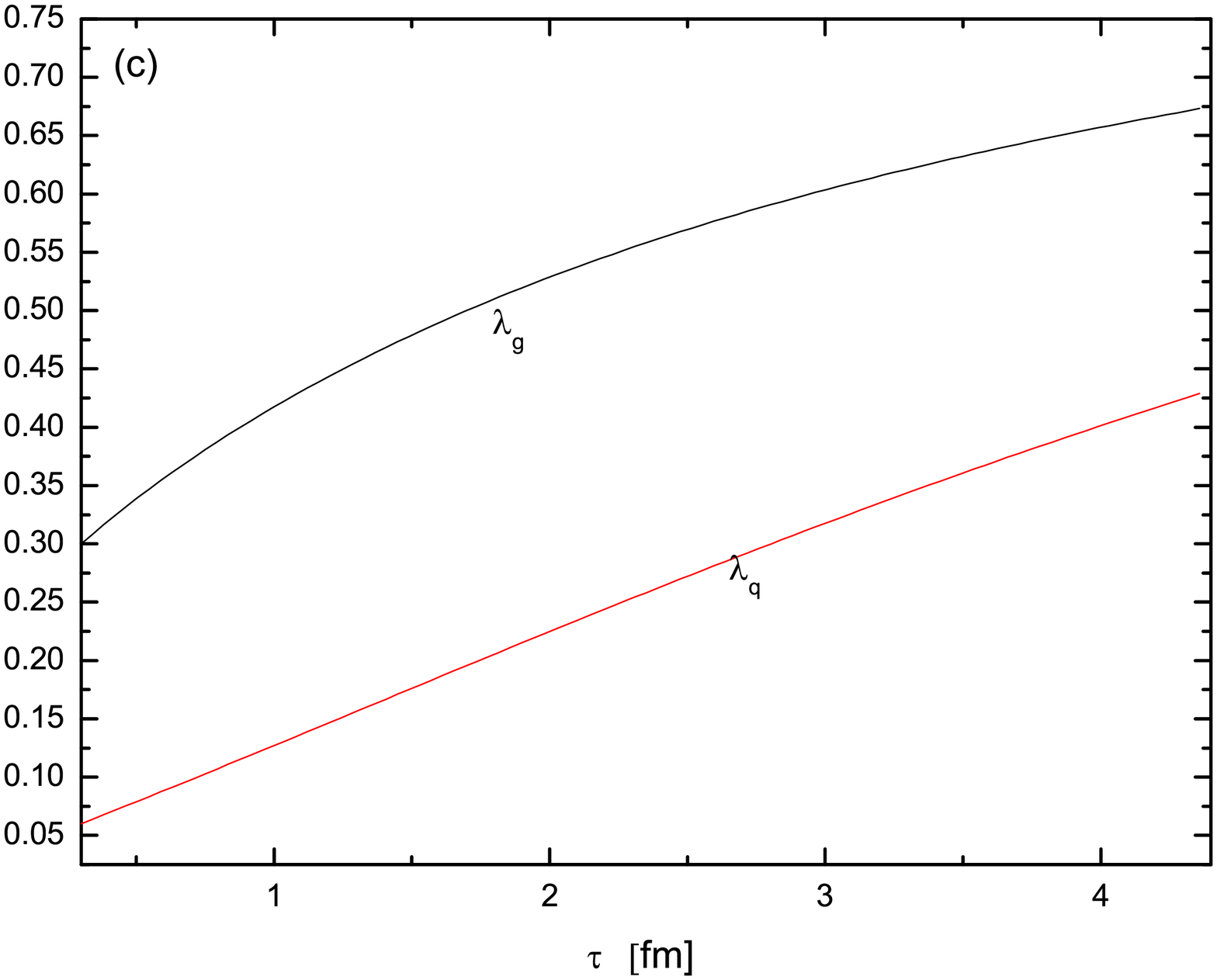}
\vspace{-20pt}\caption[a]{(a) The initial fugacity $\lambda_{g0}$
dependence of transverse energy per unit rapidity for the most
central events with different initial temperatures $T_0$=550MeV,
650MeV, 750MeV. The area between the two dot line is the the data
from RHIC that $dE_T/dy|_{y=0}=620\pm33 GeV$ for the 5\% most
central events of Au-Au collisions\cite{Adams:2004sf}.(b)Time
evolution of the temperature $T$ in chemical non-equilibrated medium
and in Bjorken expanding medium in Au+Au collisions for 200
GeV/nucleon at RHIC.(c)Time evolution of fugacity in chemical
non-equilibrated medium in Au+Au collisions for 200 GeV/nucleon at
RHIC.} \label{fig:detdydndy}\vspace{-10pt}
\end{figure}

Using the momentum distribution discussed above, we can obtain the
transverse energy per unit rapidity and parton evolution as shown in
Fig.\ref{fig:detdydndy}. It implies that the initial temperature is
$550 MeV$, $\lambda_{g0}$ is $0.3$ for the central events of Au-Au
collisions at RHIC. Using the same method, we get that the initial
temperature is $420MeV$ for Bjorken expansion($T^3\tau=T_0^3\tau_0$)
for thermal and chemical equilibrium system. The initial condition
determined is showed to be consistent with that from the particle
multiplicities.

From the evolution, the Debye screening mass, mean free path, cross
section and opacity can be obtained from the perturbative QCD at
finite temperature in a thermal equilibrated, but chemical
non-equilibrated medium\cite{Cheng}.

Since the contribution of the first order opacity is dominant, by
including the interference between the process of the rescattering
and non-rescattering, we obtain the energy loss for stimulated
emission and energy gain for thermal absorption as
\begin{eqnarray}
\Delta E_{rad}^{(1)} &=& -\frac{\alpha_s C_F E}{\pi}
\int_{\tau_0}^{\tau_0+L} d\tau \int dz \int
\frac{dk_{\perp}^2}{k_{\perp}^2} \int d^2 q_{\perp}\mid
\overline{v}(q_{\perp}) \mid^2 \frac{k_{\perp}\cdot
q_{\perp}}{(k_{\perp}-q_{\perp})^2}\nonumber \\
&&P(z)(\sigma_{gg}\rho_g+\sigma_{gq}\rho_q) <Re(1-e^{i\omega_{1}
y_{10}})> \theta(1-z),
\\
\Delta E_{abs}^{(1)} &=& \frac{\alpha_s C_F E}{\pi}
\int_{\tau_0}^{\tau_0+L} d\tau \int dz \int
\frac{dk_{\perp}^2}{k_{\perp}^2} \int d^2 q_{\perp}\mid
\overline{v}(q_{\perp}) \mid^2 \frac{k_{\perp}\cdot
q_{\perp}}{(k_{\perp}-q_{\perp})^2}\nonumber \\
 &&f_{g}(zE)(\sigma_{gg}\rho_g+\sigma_{gq}\rho_q)[P(-z)) <Re(1-e^{i\omega_{1} y_{10}})>\nonumber\\
&&-P(z)<Re(1-e^{i\omega_{1} y_{10}})>\theta(1-z)].
\end{eqnarray}
where $\omega_{1}=(k_{\perp}-q_{\perp})^2/{2\omega}$, the factor
$(1-e^{i\omega_{1} y_{10}})$ reflects the destructive interference
arising from the non-Abelian LPM effect. Averaging over the
longitudinal target profile is defined as $<\cdots>=\int dy
\rho(y)\cdots$, where $\rho(y)=2exp(-2y/L)/L$. $|{\bar v}({\bf
q}_{\perp})|^2$ is the normalized distribution of momentum transfer
from the scattering centers.

\begin{figure}
\vspace{-20pt}
\includegraphics[width=6cm]{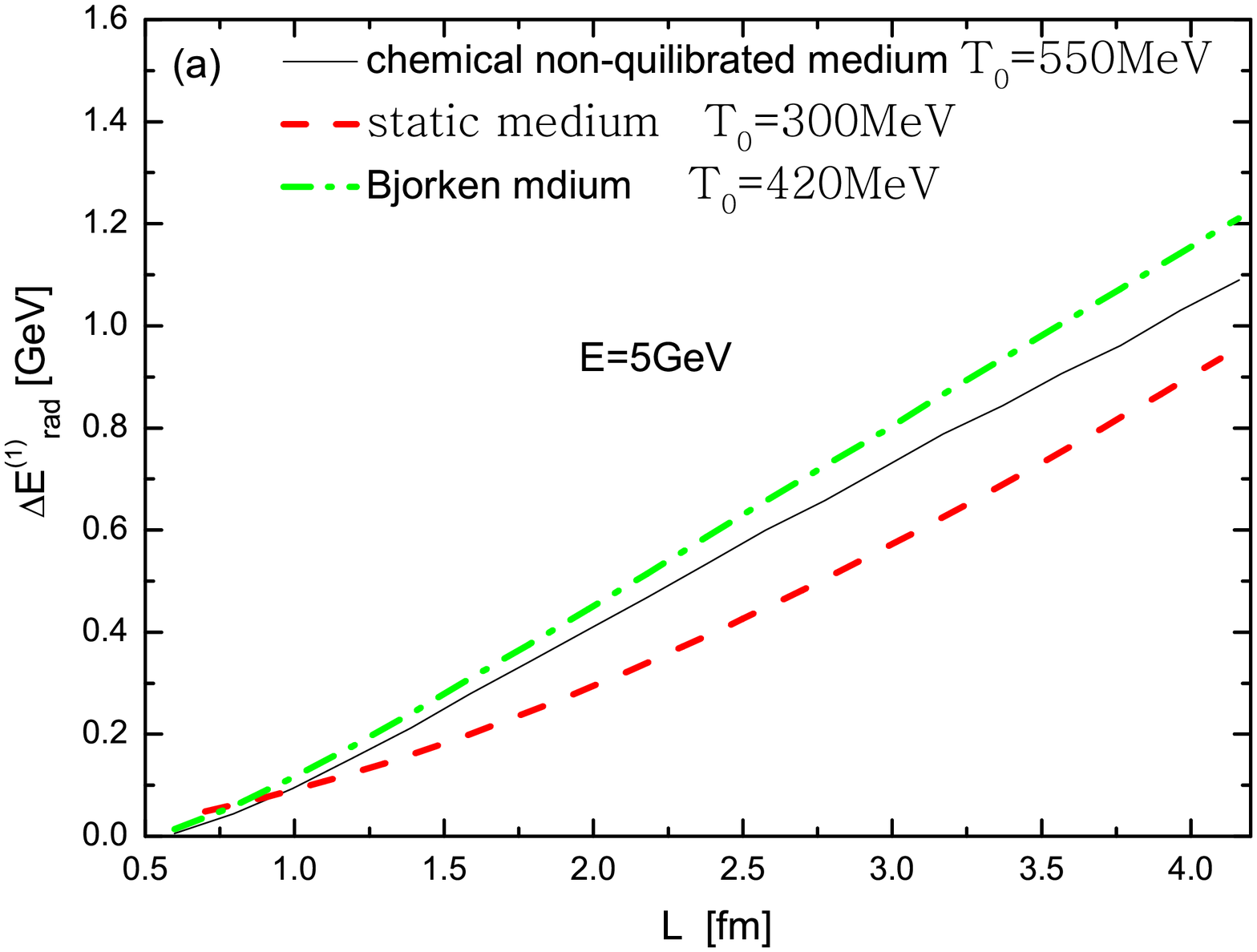}
\includegraphics[width=6.5cm]{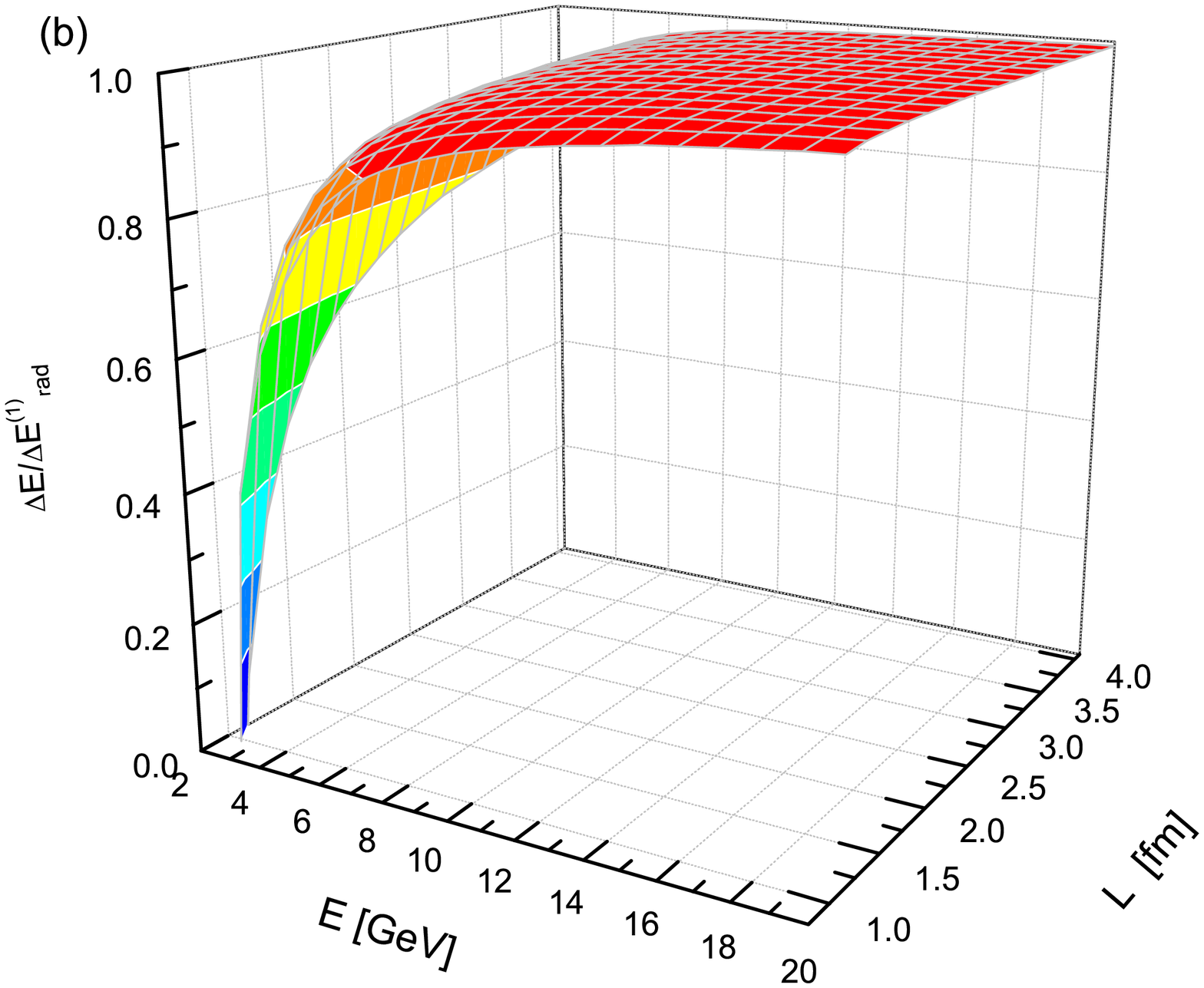}
\vspace{-10pt}\caption[a]{(a) Propagating distance dependence of the
energy loss
 to the first order
opacity. (b) The ratio of effective parton energy loss with and
without absorption as a function of parton energy $E$ and
propagating distance $L$ in the chemical non-equilibrated medium.}
\label{fig:deradcompare}\vspace{-10pt}
\end{figure}

The propagating distance dependence of the energy loss for
stimulated emission and the ratio of the calculated parton energy
loss with and without thermal absorption as functions of parton
energy value $E$ and propagating distance $L$ in a chemical
non-equilibrated medium is shown in Fig. \ref{fig:deradcompare}. It
is shown that, by taking into account the evolution of the
temperature and fugacity, the energy loss from stimulated emission
is proportional to $L$ in the chemical non-equilibrated medium and
Bjorken expanding medium rather than $L^2$-dependence on the
propagating distance in the static medium\cite{GLV}. The energy loss
in the chemical non-equilibrated medium is a bit less than that in
Bjorken expanding medium.  The energy absorption can not be
neglected at intermediate jet energies and small propagating
distance of the energetic parton in contrast with that it is
important only at intermediate jet energy in the static
medium\cite{Enke}.

In summary, we determined the initial conditions and parton
evolution and proposed the evolution effects on parton energy loss.
The evolution of the medium modifies the jet energy loss in the
intermediate energy region and affect the shape of suppression
intermediate high $P_{T}$ hadrons spectrum.

This work was supported by NSFC of China under Projects No.
10825523, No. 10635020, by MOE of China under Projects No. IRT0624,
by MOST of China under Project No. 2008CB317106; and by MOE and
SAFEA of China under Project No. PITDU-B08033.

\end{document}